\documentclass[manuscript]{emulateapj}
\usepackage{amsmath}
\usepackage{natbib}
\usepackage{hyperref}
\usepackage{xcolor}

\newcommand{\flux}{\,erg\,cm$^{-2}$\,s$^{-1}$}
\newcommand{\lum}{\,erg\,s$^{-1}$}
\newcommand{\kms}{\,km\,s$^{-1}$}
\newcommand{\cm}{\,cm$^{-2}$}
\newcommand{\nh}{$N_\mathrm{H}$}

\shorttitle{}
\shortauthors{Li et al.}

\begin{document}
\title{Discovery of a redback millisecond pulsar candidate: 3FGL J0212.1+5320}
\author{
Kwan-Lok Li\altaffilmark{1}, 
Albert K. H. Kong\altaffilmark{2}, 
Xian Hou\altaffilmark{2,3,4}, 
Jirong Mao\altaffilmark{3,4}, 
Jay Strader\altaffilmark{1}, 
Laura Chomiuk\altaffilmark{1}, 
Evangelia Tremou\altaffilmark{1}
}
\altaffiltext{1}{Department of Physics and Astronomy, Michigan State University, East Lansing, MI 48824, USA; \href{mailto:liliray@pa.msu.edu}{liliray@pa.msu.edu} (KLL)}
\altaffiltext{2}{Institute of Astronomy and Department of Physics, National Tsing Hua University, Hsinchu 30013, Taiwan}
\altaffiltext{3}{Yunnan Observatories, Chinese Academy of Sciences, Kunming, 650216, China}
\altaffiltext{4}{Key laboratory for the Structure and Evolution of Celestial Objects, Chinese Academy of Sciences, Kunming, 650216, China}

\begin{abstract}
We present a multi-wavelength study of the unidentified \textit{Fermi} object, 3FGL J0212.1+5320. Within the 95\% error ellipse, \textit{Chandra} detects a bright X-ray source {(i.e., ${F_\mathrm{{0.5-7keV}}=1.4\times10^{-12}}$\flux)}, which has a low-mass optical counterpart ($M\lesssim0.4M_\sun$ and $T\sim6000$K). A clear ellipsoidal modulation is shown in optical/infrared at 20.87 hours. The gamma-ray properties of 3FGL J0212.1+5320 are all consistent with that of a millisecond pulsar, {suggesting} that it is a $\gamma$-ray redback millisecond pulsar binary with a low-mass companion filling $\gtrapprox$ 64\% of the \textit{Roche}-lobe. {If confirmed, it will be a redback binary with one of the longest orbital periods known. }
Spectroscopic data taken in 2015 from the \textit{Lijiang} observatory show no evidence of strong emission lines, revealing {that} the accretion is currently inactive {(the rotation-powered pulsar state)}. 
{This is consistent with the low X-ray luminosities (${L_X\approx10^{32}}$\lum) and the possible X-ray modulation seen by \textit{Chandra} and \textit{Swift}. Considering that the X-ray luminosity and the {high} X-ray-to-$\gamma$-ray flux ratio {(8\%)} are both comparable} to that of the two known $\gamma$-ray transitional millisecond pulsars, {we suspect that} 3FGL J0212.1+5320 could be a {potential} target to search for future transition to the accretion active state. 
\\ 
\end{abstract}

\section{Introduction}
Progenitors of millisecond pulsars (MSPs), though not yet fully understood, are believed to be {neutron stars} in low-mass X-ray binaries (LMXBs). According to the recycling scenario \citep{1982Natur.300..728A}, {the neutron stars} are spun up through accretion from the late-type companions (if any) to ultimately evolve into MSPs. Through the so-called \textit{LMXB Case A} channel \citep{2011ASPC..447..285T}, a compact binary (i.e., orbital period $<$ 1 day) consisting of a MSP and a very low-mass companion (which was striped by the neutron star and/or partially ``evaporated'' by the energetic pulsar wind/$\gamma$-rays; \citealt{2013ApJ...775...27C}) remains at the very end phase of such an evolution, known as black widow (BW; companion mass: $<0.1M_\odot$) or redback (RB; companion mass: $\sim$0.1--0.4 $M_\odot$) binaries. A few RBs, known as transitional MSPs, have already shown remarkable transition(s) between the LMXB state and the radio pulsar state in optical, X-rays, and{/or} $\gamma$-rays {(i.e., M28I; \citealt{2013Natur.501..517P}, PSR J1023+0038; \citealt{2009Sci...324.1411A,2014ApJ...781L...3P}, and PSR J1227$-$4853; \citealt{2015ApJ...800L..12R})}, clearly indicating the close relationship between LMXBs and radio MSPs. 
BW/RBs are interesting objects, not to mention the fascinating theoretical interpretation of multi-wavelength observations for individual studies (e.g., the keV-to-GeV emission models of PSR J1023+0038 in different states; \citealt{2014ApJ...797..111L,2015ApJ...807...33P}). They also provide crucial information {on} the long-term accretion history. {In particular, BWs are the key to uncover how the companions are finally eliminated, after which isolated MSPs are formed \citep{1988Natur.334..227V}. }

As MSPs are powerful $\gamma$-ray sources with strong GeV magnetospheric radiations (e.g., from the outer gap, the slot gap, or the polar cap; \citealt{1986ApJ...300..500C,2003ApJ...588..430M,1975ApJ...196...51R}) and/or the inverse-Compton $\gamma$-ray emissions of the pulsar wind nebulae when the accretion is active \citep{2014ApJ...785..131T,2014ApJ...797..111L}, many of them should have been detected by \textit{Fermi}-LAT as a class of unidentified \textit{Fermi} object (UFO), the second-largest population detected by \textit{Fermi}-LAT \citep{2015ApJS..218...23A}. Although not all the UFOs are MSPs (in fact many of them are thought to be AGNs, the largest source class in the catalog), good BW/RBs candidates can be selected based on the $\gamma$-ray spectral curvatures and the $\gamma$-ray variabilities \citep{2012arXiv1205.3089R,2012ApJ...747L...3K,2014ApJ...794L..22K,2015ApJ...809...68H} and confirmed their pulsar natures by detecting the radio/$\gamma$-ray pulsations. Thanks to the \textit{Fermi} Pulsar Search Consortium (PSC), a great success has been achieved in discovering new pulsars through ``blind'' searches {for} coherent pulsations in radio and $\gamma$-rays \citep{2012arXiv1205.3089R}, and {the known BW and RB populations have been greatly extended in recent years. }

{Alternatively, multi-wavelength studies of UFOs are the secondary way to search for BW/RBs MSP candidates. }In most of the cases, X-ray follow-ups are the key to narrow down the source location, allowing identification of the optical counterparts. 
{Once the optical counterpart is identified}, time-series optical observations can test the BW/RB identity by searching for the orbital modulation on timescale of hours produced by pulsar irradiation on the companion and/or ellipsoidal variation. 
Through this multi-wavelength technique, several UFOs, for examples, 2FGL J1311.7$-$3429/PSR J1311$-$3430 \citep{2012Sci...338.1314P}, 1FGL J1417.7$-$4407/PSR J1417$-$4402 (not a canonical BW/RB system; \citealt{2015ApJ...804L..12S,2016ApJ...820....6C}), and 1FGL J2339.7$-$0531/PSR J2339$-$0533 \citep{2012ApJ...747L...3K,2015ApJ...807...18P} have been identified as MSP binaries and some of them have been {confirmed by the detection of millisecond radio/$\gamma$-ray pulsations}, proving the validity of the method. 

In this paper, we report the discovery of a $\gamma$-ray-emitting {RB candidate}, 3FGL J0212.1+5320. In the following sections, we present multi-wavelength studies using the optical imaging/spectroscopic data from the \textit{Lijiang} \citep{2015RAA....15..918F}, \textit{Lulin}, and Michigan State University (MSU) observatories, the \textit{Chandra} X-ray data, and the \textit{Fermi}-LAT third source catalog (3FGL; \citealt{2015ApJS..218...23A}). Discussions will be given in the last section. 

\section{{The Gamma-ray Properties in 3FGL}}
3FGL J0212.1+5320 is an unidentified bright $\gamma$-ray source (i.e., $F_\gamma=(1.71\pm0.16)\times10^{-11}$\flux\ in 0.1--100~GeV, {which is top 15\% among the sources in 3FGL; \citealt{2015ApJS..218...23A}}) that was first detected by \textit{Fermi}-LAT in $\gamma$-rays in the \textit{Fermi}-LAT first source catalog (1FGL; \citealt{2010ApJS..188..405A}). It also later appears in 3FGL with a detection significance of $25\sigma$. 

{Based on the second \textit{Fermi}-LAT pulsar catalog \citep{2013ApJS..208...17A}}, the $\gamma$-ray properties of pulsars can be characterized by a low source variability and a curved $\gamma$-ray spectral shape. Although they are not necessary conditions, 3FGL J0212.1+5320 fulfils both of the criteria (Table \ref{tab:rbs}), suggesting its possible pulsar nature in $\gamma$-rays. 
Similar to many other $\gamma$-ray pulsars that have seen stable in $\gamma$-rays over years \citep{2010ApJS..188..405A}, 3FGL J0212.1+5320 can also be considered as a steady source with a small 3FGL variability index of 51.47 (i.e., for a source with a variability index larger than 72.44, there is a less than 1\% chance of being a steady source; {\citealt{2015ApJS..218...23A}}). In addition, the $\gamma$-ray spectrum of 3FGL J0212.1+5320 is probably more than a single power-law but rather with an extra curvature component (e.g., an exponential cut-off) as the spectral curve significance is 6.3$\sigma$ in 3FGL, which is also another common feature among the pulsars detected in 3FGL \citep{2015ApJS..218...23A}. 
In fact, \cite{2016ApJ...820....8S} and \cite{2016ApJ...825...69M} have found that 3FGL J0212.1+5320 is a strong MSP candidate{, using statistical and machine learning techniques. }
\section{{\textit{Swift} and \textit{Chandra} X-ray Observations}}
\label{sec:xray}
As one of the survey targets in the \textit{Swift}/XRT survey of \textit{Fermi} unassociated sources \citep{2013ApJS..207...28S}, 3FGL J0212.1+5320 has been observed twice by \textit{Swift}/XRT in October 2010 (the observations are separated by 3 days with a total exposure time of 4.5~ks). Within the 95\% 3FGL error ellipse, a bright X-ray counterpart was detected and listed as 1SXPS J021210.6+532136 in the \textit{Swift}/XRT point source catalog (1SXPS; \citealt{2014ApJS..210....8E}). According to 1SXPS, the source is located at $\alpha\mathrm{(J2000)}=02^\mathrm{h}12^\mathrm{m}10\fs62$, $\delta\mathrm{(J2000)}=+53\arcdeg 21\arcmin 36\farcs8$ (90\% positional uncertainty: 3.8\arcsec) with a mean count rate of $(2.26\pm0.26)\times 10^{-2}$~ct~s$^{-1}$. 
A moderate flux variability is seen between the two observations from $(2.61\pm0.32)\times10^{-2}$~ct~s$^{-1}$ to $(1.31\pm0.41)\times10^{-2}$~ct~s$^{-1}$ in 3 days (equivalent to a $2.9\sigma$ change). 
The X-ray spectrum could be described by an absorbed power-law of \nh\ $=1.4^{+2.8}_{-1.4}\times10^{21}$\cm\ (the Galactic column density \nh\ $=1.5\times10^{21}$\cm; \citealt{2005A&A...440..775K}) and $\Gamma_X=1.0^{+0.5}_{-0.4}$ with an unabsorbed flux of $F_\mathrm{0.3-10keV}=1.6^{+0.5}_{-0.3}\times10^{-12}$\flux\ ($W$-$stat=57.98$ and $\chi^2=63.02$; $dof=78$). Alternatively, the spectrum could be fitted with an \texttt{APEC} thermal plasma model, however, with an extremely high and poorly constrained plasma temperature (i.e., $kT\sim100$~keV). As the best-fit temperature is just too high to be physical, we do not further {consider the \texttt{APEC} model} in the following analyses. 

\begin{figure}
\centering
\includegraphics[width=85mm]{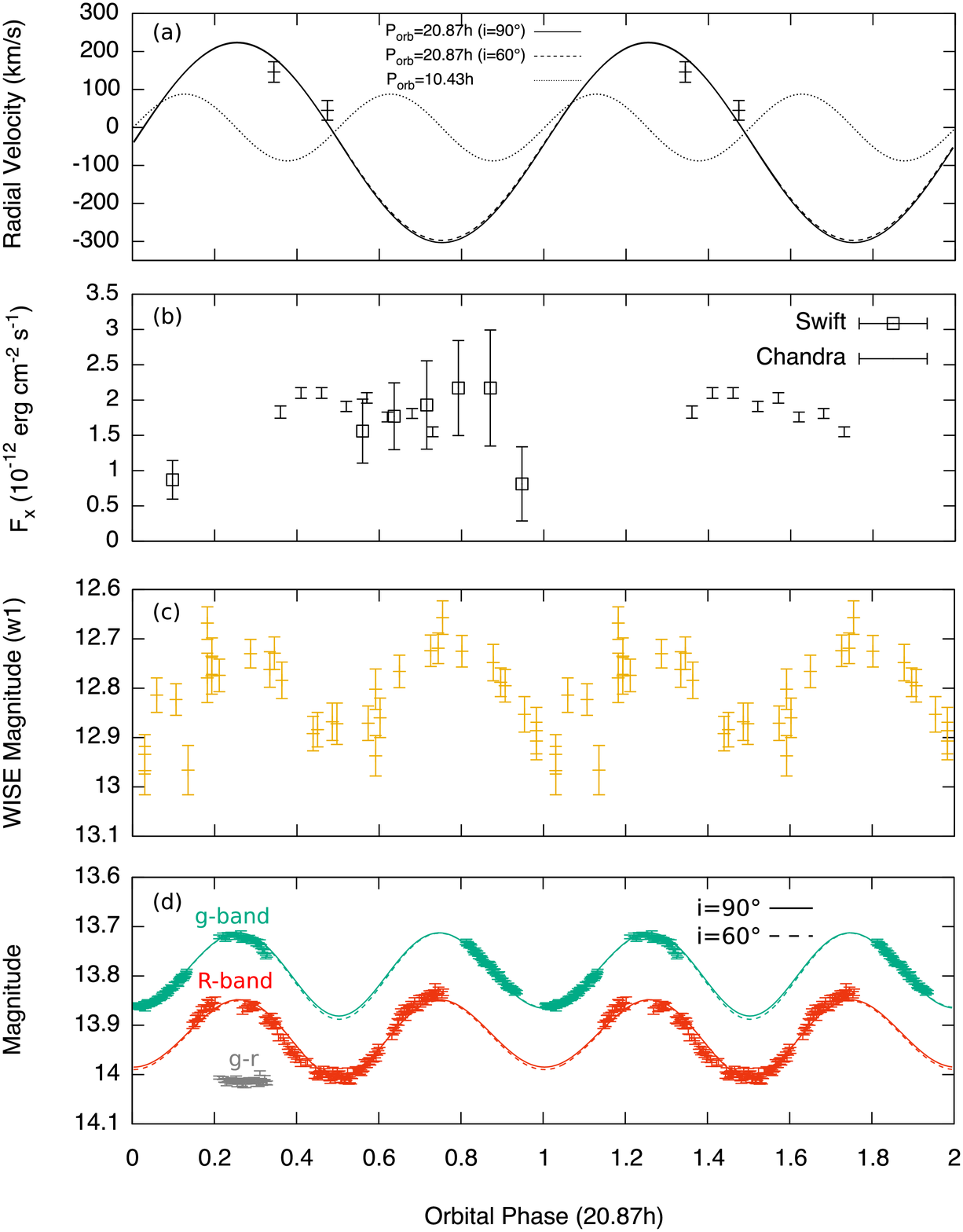}
\caption{The plots show {several physical quantities versus orbital phase} ($P_\mathrm{orb}=20.8698$~h), including (a) the radial velocities with the ELC models at 20.87h (solid line for $i=90\arcdeg$ and dashed line for $i=60\arcdeg$) and a 10.43h model curve (with an arbitrary amplitude; dotted line) projected on the 20.87h orbital phase for comparison, (b) the X-ray flux (the \textit{Swift} data is only shown in the first cycle for a clear view of the \textit{Chandra} data in the second cycle), (c) the WISE $w1$-band data, and (d) the $g$- and $R$-band data tentatively calibrated with the UCAC4 Catalog \citep{2013AJ....145...44Z} and the extinction of $A_v=0.4992$~mag \citep{2011ApJ...737..103S} with the ELC models and the ($g$-$r$) with an arbitrary offset. Two cycles are shown for clarity. 
\\
}
\label{fig:phase_lc}
\end{figure}

\textit{Chandra} has also observed the field of view once with ACIS for 30~ks in 2013 August (Obs ID: 14814; PI: Saz Parkinson) and 1SXPS J021210.6+532136 is clearly detected at $\alpha\mathrm{(J2000)}=02^\mathrm{h}12^\mathrm{m}10\fs50$, $\delta\mathrm{(J2000)}=+53\arcdeg 21\arcmin 38\farcs9$ (90\% positional uncertainty: 0.8\arcsec) with a net count rate of $(9.03\pm0.17)\times10^{-2}$~ct~s$^{-1}$ (0.5--7~keV). With a total number of 2685 photon counts, we binned the data to have at least 20 counts per bin and fitted the binned spectrum with an absorbed power-law. The best-fit parameters are \nh $=(1.4\pm0.5)\times10^{21}$\cm, $\Gamma_X=1.3\pm0.1$, and $F_\mathrm{0.5-7keV}=(1.35\pm0.06)\times10^{-12}$\flux\ (or $F_\mathrm{0.3-10keV}=(1.89\pm0.08)\times10^{-12}$\flux; $\chi^2=98.78$ and $dof=105$), which are all consistent with that extracted from the \textit{Swift}/XRT data {and the \textit{Chandra} spectral fitting by \cite{2016ApJ...820....8S}}. To examine the short term variability seen by \textit{Swift}/XRT, we extracted a 4000-sec bin lightcurve with the \textit{Chandra}/ACIS data and a flux variability on an hourly timescale is clearly shown (Figure \ref{fig:phase_lc}). To quantify the variability significance, we computed the $\chi^2$ value of the 8 data bins with a flat lightcurve model, which is $\chi^2=24.39$ ($dof=7$), indicating that there is only {a 0.1\% chance} that the variability is produced by random fluctuation. 

\begin{figure*}
\centering
\includegraphics[width=150mm]{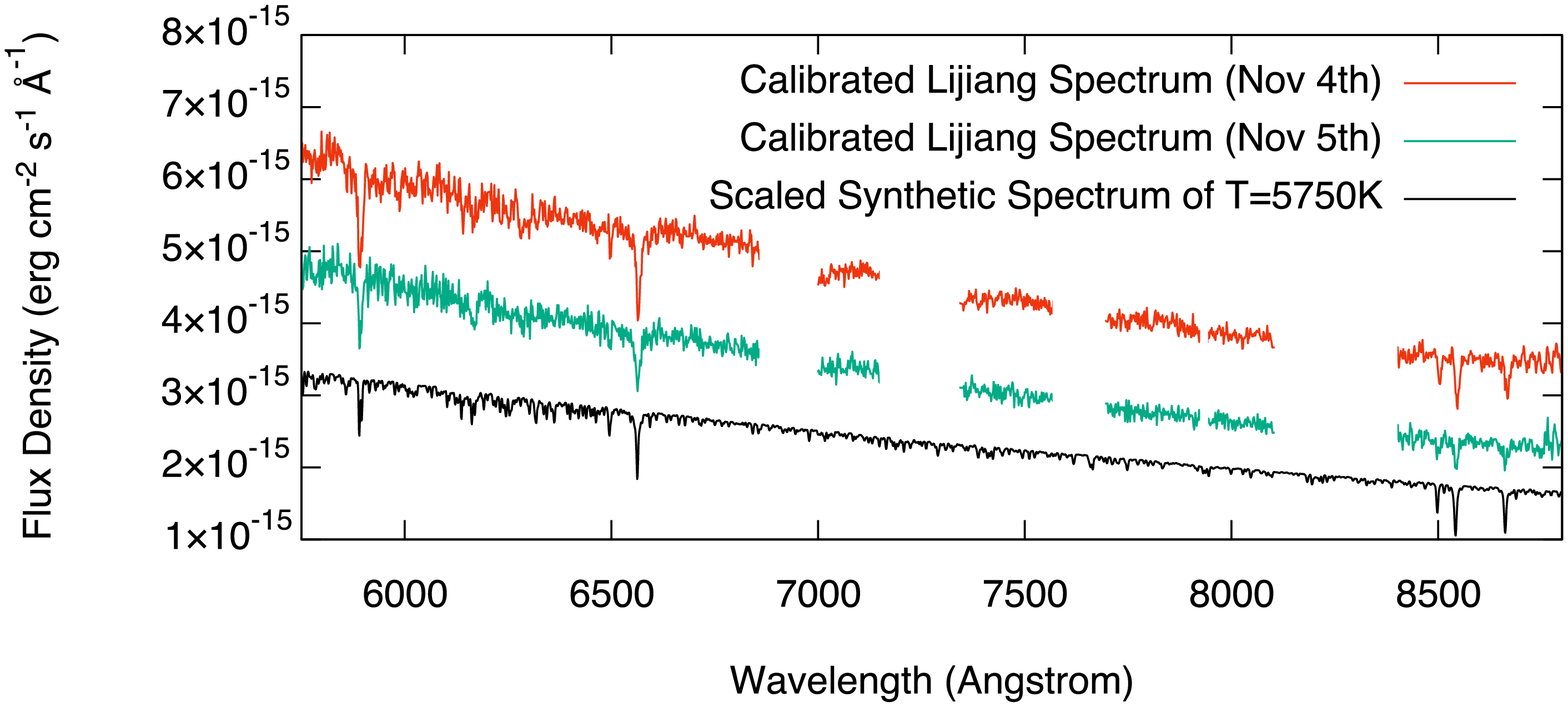}
\caption{From top to bottom, the curves are the two \textit{Lijiang} spectra of USNO-B1.0 1433-0078846 taken on November 4th and 5th, and the model spectrum of $T=5750$~K from the Munari synthetic spectral library. There are four gaps present on the \textit{Lijiang} spectra due to the removal of the telluric lines. 
\\
}
\label{fig:opt_spec}
\end{figure*}

\section{Optical data}
At the \textit{Chandra} X-ray position, we found a bright optical counterpart ($R=14.23$~mag) in the USNO-B1.0 catalog \citep{2003AJ....125..984M}, USNO-B1.0 1433-0078846, with an offset of 0.2\arcsec. The same source is also detected in the 2MASS \citep{2006AJ....131.1163S} and WISE \citep{2010AJ....140.1868W} catalogs. Using {the multi-epoch} photometry table of WISE\footnote{\url{http://irsa.ipac.caltech.edu/Missions/wise.html}}, a variability of 0.2--0.3~mag is clearly seen in the $w1$ band data of 33 epochs taken in 2010 February and August. The modulation is likely periodic with a period of $\sim$10--20 hours (see Figure \ref{fig:phase_lc}c for the modulation, although the phase light curve was folded at 20.87 hours). 

\subsection{Imaging from the MSU and \textit{Lulin} Observatories}
A monitoring campaign with the 0.6-m telescope in the MSU observatory and the 1-m telescope in the \textit{Lulin} observatory has been carried out from 2015 October to 2016 January to investigate the $\sim$10--20~h modulation seen in WISE. We observed the source for 3 consecutive nights from October 10 to October 12 with the 0.6-m telescope in the $R$-band (200/300 sec for each frame, depending on the weather) and with the 1-m telescope in the SDSS $r$- and $g$-bands for 3 other nights (i.e., November 8/9 and January 9; only $g$-band images were taken on the first two nights and both $r$- and $g$-band were taken by turns on the last night; 60/120 sec for the $r$/$g$-band images, respectively). 

\subsection{Spectroscopy from the \textit{Lijiang} Observatory}
\label{sec:lijiang}
Two 1200-sec {medium-resolution} optical spectra ($5750$--$8800\mathrm{\AA}$) were taken on 2015 November 4 and 5 with the 2.4-m telescope at the \textit{Lijiang} observatory. After (i) the standard reduction processes with the \texttt{IRAF} package \texttt{ONEDSPEC}, (ii) a flux calibration with the standard star BD+28$\arcdeg$~4211 \citep{1990AJ.....99.1621O}, and (iii) an extinction correction with $A_v=0.4992$~mag (\citealt{2011ApJ...737..103S}; which is roughly consistent with the \nh\ value estimated by \textit{Chandra}) and the \textit{Cardelli} extinction law \citep{1989ApJ...345..245C}, the calibrated data show spectral shapes comparable to that of a low-mass star (Figure \ref{fig:opt_spec}) without any accretion features. After matching the data with the synthetic spectra from the Munari online library\footnote{\url{http://archives.pd.astro.it/2500-10500/}} (\citealt{2005A&A...442.1127M}; a solar metallicity of [M/H] $=0$ and a typical RB rotational broadening of $V=100$\kms\ are assumed), we found that the spectra can be best described by $T=5750$~K and $\log g=4.5$ (Figure \ref{fig:opt_spec}), of which the stellar properties are very close to the $M\approx0.4M_\sun$ low-mass companion of the RB PSR J2129$-$0429 \citep{2016ApJ...816...74B}. Therefore, we tentatively assume the secondary star of 3FGL J0212.1+5320 to be around $M\sim0.4M_\sun$. 

\section{Detailed Timing analyses}

\subsection{Orbital Period Determination}
After applying the standard data reduction procedures by \texttt{IRAF} on the optical imaging data and removing some bad frames due to bad tracking or bad weather, we used a differential photometry technique to study the optical modulation, which shows a clear sinusoidal shape in all bands (Figure \ref{fig:phase_lc}d). We fitted all the data (including the WISE data; all are heliocentric corrected) simultaneously with sinusoidal functions with common period and phases, but different amplitudes and baselines for each data set. The best-fit period is 10.43479(7) hours (corresponding to the pulsar irradiation case) or 20.8698(1) hours (the ellipsoidal variation case) with the flux minimum epoch at HJD 2457305.5551(4) (the phase zero of Figure \ref{fig:phase_lc} and the following timing analyses). 
It is worth noting that the data used span over 5 years of time (i.e., from 2010 to 2015), which leads to a very high accuracy of the best-fit period. 
The best-fit amplitudes of the bands are roughly {consistent with each other} within {a} largest offset of 0.02~mag (i.e., $a_{w1}=0.09\pm0.02$~mag, $a_R=0.0845\pm0.0009$~mag, $a_g=0.0731\pm0.0004$~mag, and $a_r=0.092\pm0.007$). In particular, the simultaneous $r$- and $g$-band data taken by \textit{Lulin} on January 9 do not show any clear color evolving trend during the phase interval of $\phi_{10}=0.42-0.66$ at $P=10.43$h (or $\phi_{20}=0.21-0.33$ at $P=20.87$h; Figure \ref{fig:phase_lc}d), suggesting that there is likely no strong orbital color variability. This indicates that the pulsar irradiation effect on the companion is very limited and thus the modulation is probably caused by ellipsoidal variation. 

\begin{figure}
\centering
\includegraphics[width=85mm]{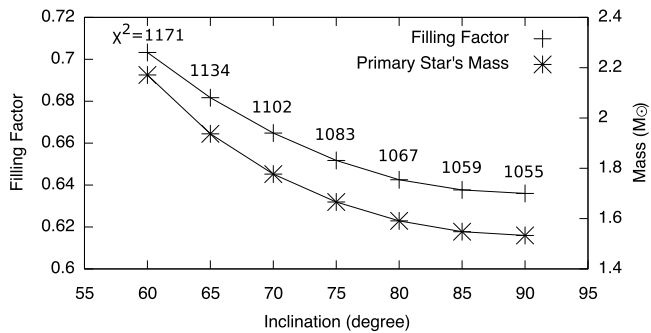}
\caption{The above two curves show the best-fit filling factors of the secondary star and the inferred masses of the primary star {versus} the binary inclinations from $60\arcdeg$ to $90\arcdeg$ with the corresponding best-fit $\chi^2$ values. \\
}
\label{fig:elc_fit}
\end{figure}

\subsection{Radial Velocity Measurement}
Following the method described in \cite{2016ApJ...816...74B}, we first removed the telluric lines of the \textit{Lijiang} spectra by
omitting bands of 6860--7000$\mathrm{\AA}$, 7570--7700$\mathrm{\AA}$, 7150--7350$\mathrm{\AA}$, and 8100--8400$\mathrm{\AA}$. Using the \texttt{RVSAO} Package of \texttt{IRAF}, we used the task \texttt{xcsao} to calculate the barycentric corrected radial velocities by cross-correlating the spectral data with the $T=5750$~K synthetic spectrum (all the spectra involved are automatically normalized during the cross-correlation process). Both the spectra were found to be red-shifted with radial velocities of $136\pm19$\kms\ (Nov 4) and $31\pm17$\kms (Nov 5). By applying the 20.87h (or 10.43h) ephemeris, the orbital phases of the radial velocities are $\phi_{20}=0.34$ (or $\phi_{10}=0.69$) and $\phi_{20}=0.47$ (or $\phi_{10}=0.95$), respectively. For the pulsar irradiation case (i.e., the orbital period is 10.43 hours), the companion should be moving from the behind of the pulsar to the front in the orbital interval of $\phi_{10}=0.5-1$ (i.e., $\phi_{20}=0.25-0.5$ in Figure \ref{fig:phase_lc}a), during which the lowest radial velocity (RV) occurs at $\phi_{10}=0.75$ (i.e., $\phi_{20}=0.375$). Therefore, the radial velocity at $\phi_{10}=0.95$ (i.e., $\phi_{20}=0.47$) should be higher than that at $\phi_{10}=0.69$ (i.e., $\phi_{20}=0.34$). However, the result shows differently, indicating the invalidity of the irradiation case (see Figure \ref{fig:phase_lc}a for a more clear demonstration). On the contrary, the observed radial velocities can be naturally explained in the case of ellipsoidal variation if the orbital phase zero is defined as the inferior conjunction (i.e., the companion is between the pulsar and the observer; Figure \ref{fig:phase_lc}a). 

\subsection{ELC fitting}
We used the \textit{Eclipsing Light Curve} (ELC) code (Version 3; \citealt{2000A&A...364..265O}) to model the optical lightcurves (i.e., $R$- and $g$-bands) obtained from the MSU and \textit{Lulin} observatories for a deeper understanding of the interacting binary. For the $R$-band data, we omitted the short $r$-band lightcurve (i.e., 2.4 hours) obtained from \textit{Lulin} to prevent extra systematic uncertainties originated from the cross-calibrations between different filter systems (i.e., $r$- and $R$-bands) and instruments. 
As ELC is capable of fitting RV, we also considered the two radial velocities to have a better constraint on the fitting result, despite the limited data quantity/quality. We also allowed a tiny phase shift between the phase-folded lightcurves and the models to further calibrate for the epoch of the inferior conjunction (i.e., the phase zero of the ELC models). 

By (i) using the orbital period of $P_\mathrm{orb}=20.8698$h, (ii) assuming the effective temperature of the companion $T_\mathrm{eff}=5750$K (it was not well determined because the ELC fit is insensitive to the companion temperature as ELC fits the normalized light curves) and the mass of the secondary star is $m_2\sim0.4M_\sun$ (by setting $m_2=0.3-0.5M_\sun$), (iii) disabling the radiation heating effect, (iv) adopting the linear limb-darkening law \citep{1993AJ....106.2096V} with a coefficient of $\kappa=0.6483$ \citep{2010A&A...510A..21S}, and (v) setting a circular orbit (i.e. $e=0$), we fitted the lightcurves by varying four binary parameters, which are the binary inclination ($i$), the mass ratio ($q=m_1/m_2$, where $m_1$ is the pulsar mass), the orbital separation ($a$), and the \textit{Roche}-lobe filling factor ($\beta$, the ratio of the volume-averaged radii of the companion star and the \textit{Roche}-lobe; \citealt{1984ARA&A..22..537J}). With the built-in optimizer \texttt{gridELC}, we searched for the best-fit solution by minimizing the $\chi^2$ value and the least reduced chi-square of $\chi^2_\nu=3.1$ ($dof=344$) was found at $i=90\arcdeg$, $q=6.8$, $a=4.6R_\sun$, and $\beta=0.64$ ($m_1=1.5M_\sun$ and $m_2=0.2M_\sun$ are inferred). It is not a good fit statistically and the extreme inclination at the upper-bound may imply that the fit did not converge\footnote{We once considered turning on the radiation heating effect to improve the fit. However, the flux overestimation in the valley at $\phi\sim0.5$ is the main cause of the bad fitting. The radiation heating effect will even increase the predicted flux there to worsen the fit. }. 
In addition, we found that the data can be fitted fairly well even if a fixed inclination angle of different value is used (see the similarity between the best-fit models at different inclinations in Figure \ref{fig:phase_lc}d). 
Therefore, instead of estimating the parameter uncertainties, we obtain and discuss the best-fit parameter sets at different inclinations from $i=90\arcdeg$ to $i=60\arcdeg$ with a step size of $5\arcdeg$. To elaborate the choice of $i=60\arcdeg$, it was chosen based on the $\chi^2$ values of the RVs on the ELC fits ($\chi_\mathrm{RV}^2$). Despite the complexity of the optical emission revealed by the bad ELC fit, the RV data in principle would not be affected making $\chi_\mathrm{RV}^2$ a useful indicator to test the model validity. In this case, we chose a criteria of $\chi_\mathrm{RV}^2<3.84$ (95\% c.l. for $dof=1$) to reject all other steps of $i<60\arcdeg$ with {large} $\chi_\mathrm{RV}^2$. 
We note that the selection heavily depends on the weighting on the RV data in the ELC fit (i.e., no weighting applied here) and therefore the rejection does not imply that the inclination has to be $i>60\arcdeg$. The selection simply indicates that the unweighted best-fit models with $i<60\arcdeg$ are inconsistent with the RV data and thus no discussion will be given on those fits. 

Figure \ref{fig:elc_fit} shows the best-fit \textit{Roche}-lobe filling factors and the inferred pulsar masses for $i=60\arcdeg-90\arcdeg$. As expected, the best-fit filling factor decreases with the inclination (from 0.70 to 0.64). {All the best-fit results lead the primary star's mass to the range of $m_1=1.5-2.2M_\sun$ (and $m_2\approx0.2M_\sun$), which is consistent with that of a pulsar.} Certainly, the errors of the best-fit pulsar masses could be large (e.g., the uncertain companion mass as one of the major sources of error). 
Also, we found that the ELC fits are not robust. For instance, if we {remove the constraints on} the RV curve (i.e., the two \textit{Lijiang} data points) and the companion mass (i.e., $m_2=0.3-0.5M_\sun$), the best-fit solution of $i=90\arcdeg$ changes to $q=7.7$, $a=5.0R_\sun$, and $\beta=0.63$, with which $m_1=2.0M_\sun$ and $m_2=0.3M_\sun$ are inferred (cf. Figure \ref{fig:elc_fit}). Even two RV data points and {a weak constraint} on the companion mass are sufficient to significantly affect the fitting result. Therefore, we conclude that the best-fit parameters and the inferred masses are merely indicative. {Detailed modelling} (e.g., by adding hot/cool spots on the companion) as well as high-quality photometry sets and spectroscopic data of a complete orbit are required to place a further constraint on the pulsar mass. More imaging and spectroscopic observations are being planned to probe the system in the near-future. 

\setlength{\tabcolsep}{3pt}
\begin{table*}
\centering 
\caption{\footnotesize{X/$\gamma$-ray properties of some known RBs in the pulsar state and 3FGL J0212.1+5320}}
\begin{tabular}{@{}lcccccc}
\hline
Name & Spectral Curvature\footnotemark[1] & Variability\footnotemark[2] & $F_\mathrm{0.1-100GeV}$ & $\Gamma_X$ & $F_\mathrm{0.5-7keV}$ & $F_X/F_\gamma$\\
& ($\gamma$-ray; $\sigma$) & ($\gamma$-ray ) & ($10^{-11}$\flux) & & ($10^{-13}$\flux) & \\
\hline
\multicolumn{7}{c}{(Generic $\gamma$-ray emitting RBs)}\\
PSR J2129$-$0429 & 3.7 & 60.3 & 1.1 & 1.3 & 0.11 & 0.10\% \\
PSR J2339$-$0533 & 8.7 & 40.1 & 3.0 & 1.4	& 1.4 & 0.48\%\\
PSR J1628$-$3205 & 5.5 & 50.5 & 1.2 & \multicolumn{2}{l}{(no X-ray detection)} & $<$1.1\%\\
PSR J1048+2339 & 2.5 & 49.7 & 0.7 & \multicolumn{2}{l}{(no X-ray detection)} & $<$1.9\%\\
\hline
\multicolumn{7}{c}{(Prospective tMSP Candidates in the Pulsar State)}\\
PSR J2215+5135 & 6.8 & 56.9 & 1.4 & 1.8 & 1.0 & 0.74\%\\
PSR J1723$-$2837 & 3.3 & 55.7 & 1.8 & 0.9 & 24 & 13\%\\
\hline
\multicolumn{7}{c}{(\textit{Fermi}-detected tMSPs in the Pulsar State)}\\
PSR J1227$-$4853 & \nodata & \nodata & 0.4 & 1.2 & 4.6 & 13\%\\
PSR J1023+0038 & \nodata & \nodata & 0.1 & 0.9 & 4.7 & 37\%\\
\hline
\multicolumn{7}{c}{(Our Target)}\\
{3FGL J0212.1+5320} & {6.3} & {51.5} & {1.7} & {1.3} & {14} & {7.9\%} \\
\hline
\end{tabular}
\flushleft
References: 3FGL \citep{2015ApJS..218...23A}; \cite{2010ApJ...724L.207T,2012ApJ...747L...3K,2014ApJ...795...72L,2015ApJ...801L..27H,2015ApJ...808...17X,2016ApJ...823..105D}
\footnotetext{3FGL curvature index: significance of the fit improvement between power-law and either LogParabola or PLExpCutoff spectrum type. }
\footnotetext{3FGL variability index: a value greater than 72.44 indicates there is a less than 1\% chance of being a steady source. }
\label{tab:rbs}
\end{table*}

\subsection{Possible X-ray Orbital Modulation}
As mentioned in $\S$\ref{sec:xray}, there is a significant variability seen in both \textit{Swift}/XRT and \textit{Chandra} data, which is possibly induced by the X-ray orbital modulation. We thus folded the lightcurve with the 20.87h timing solution after converting the \textit{Chandra} X-ray flux into the \textit{Swift}/XRT band (i.e., 0.3--10~keV) and performing a barycentric correction to the data. Although the folded X-ray lightcurve does not cover a full orbital cycle, the X-ray variation is likely periodic with an X-ray minimum around the inferior conjunction (Figure \ref{fig:phase_lc}b). A similar {phenomenon has been previously seen} in the RB PSR J1023+0038 {\citep{2011ApJ...742...97B,2014ApJ...791...77T,2014ApJ...797..111L}}. From the \textit{Chandra} data bins, the X-ray maximum occurs around the superior conjunction (i.e., $\phi_{20} \sim 0.5$; observer-pulsar-companion), although the \textit{Swift} data favours the flux maximum around $\phi_{20} > 0.5$. 

\section{Discussion and Conclusion}
We presented a multi-wavelength study of 3FGL J0212.1+5320 and found that {a RB MSP} binary as its physical nature can naturally explain the entire data set. 
The X/$\gamma$-ray spectral properties and the hourly-timescale orbital period are very similar to that of many known RBs (Table \ref{tab:rbs}), revealing the first hint of 3FGL J0212.1+5320 as a RB candidate. {The inferred primary star's masses from the best-fit ELC models are $1.5-2.2\,M_\sun$ that are consistent with that of a neutron star, though they are only indicative estimates.} An hourly variability is seen in the \textit{Swift}/\textit{Chandra} joint lightcurve and it could be an orbital modulation, however, uncertainly. If the modulation is genuine, it could be caused by an intrabinary shock emission, {through \textit{Doppler} boosting with a pulsar-wrapping shock geometry \citep{2014ApJ...797..111L} or partial occultation by the companion \citep{2011ApJ...742...97B}}. All the observational evidence is pointing to the conclusion of 3FGL J0212.1+5320 as a newly-discovered RB system. 

{A bright optical counterpart (could be one of the brightest known for RBs) has been identified with a clear orbital modulation at 20.87 hours}. We do not see an obvious non-uniform radiation heating to contribute {to} the orbital modulation and therefore the companion is probably not completely tidally locked. This may imply 3FGL J0212.1+5320 as a very young MSP system. According to \cite{1977A&A....57..383Z}, the synchronization timescale of such a close binary is approximately $t_\mathrm{sync}\sim10^4\,((1+q_i)/2)^2(P_i/\mathrm{1\,day})^4$~years (equation 6.1 of \citealt{1977A&A....57..383Z}), where $q_i$ and $P_i$ are the initial mass ratio and orbital period, respectively\footnote{The equation presented here is slightly different from the one in \citealt{1977A&A....57..383Z} because of the different definitions of the mass ratios. }. 
Assuming an initial mass ratio of $q_i=2.8$ (i.e., $m_{1,i}=1.4M_\sun$ and $m_{2,i}=0.5M_\sun$)\footnote{The initial masses are both poorly known due to the highly uncertain accretion and ablation processes, and thus the {values} are merely estimated within reasonable ranges. }, 
$P_i\approx13$d gives $t_\mathrm{sync}\gtrsim10^9$ years and $P_i\approx4$d gives $t_\mathrm{sync}\gtrsim10^7$ years. 
We took the calculated timescales for 3FGL J0212.1+5320 as lower limits because the orbital widening by the ablation from the pulsar \citep{2013ApJ...775...27C}, that would extend the synchronization process, was not considered in Zahn's work. 
In the case of $t_\mathrm{sync}\gtrsim10^7$ years, the initial orbital period is actually close to the estimated value of PSR J2129$-$0429 (i.e., $P_i\approx2.5$d; \citealt{2016ApJ...816...74B}), which has a long orbital period of $P=15.2$h, comparable to 3FGL J0212.1+5320's. Obviously, a young age of 3FGL J0212.1+5320 (i.e., in the order of 10 Myr) would be a self-consistent explanation for the data. In fact, $\sim10$ Myr old MSPs are rare but not impossible. For example, PSR J1823$-$3021A, one of the youngest MSPs known, has a characteristic age of 25 Myr \citep{2011Sci...334.1107F}. {Searching for the radio/X-/$\gamma$-ray pulsations of 3FGL J0212.1+5320 and computing the characteristic age would be useful to investigate the speculation. }

Despite no heating effect seen, it is still highly likely that the companion is uniformly irradiated by the X/$\gamma$-rays from the pulsar, {resulting in a higher} surface temperature than a $\sim0.4\,M_\sun$ star should have. As the companion mass is no longer the only dominant factor to determine the surface temperature, the assumption of $m_2\sim0.4M_\sun$ (see \S \ref{sec:lijiang})
could be overestimated. Considering the fact that all the fitting results indicate a lighter $m_2$, $m_2\lesssim0.4M_\sun$ would be more reasonable. 

As the companion has a temperature {close to that of the Sun}, it is convenient to use the solar $R$-band absolute magnitude (i.e., $R=4.42$~mag; \citealt{1998gaas.book.....B}) to infer the distance of 3FGL J0212.1+5320. From the ELC model fits, the size of the companion is about $R_c\approx1\,R_\sun$. After a proper scaling, the inferred distance is about $d\approx0.8$~kpc leading to an X-ray luminosity of $L_X\approx10^{32}$\lum, which is relatively high among the known X-ray RBs in the \textit{pulsar state} (when radio pulsations can be detected and $L_X\sim10^{31}-4\times10^{32}$\lum; \citealt{2014ApJ...795...72L}). {Since} a high X-ray luminosity (i.e., $L_X\gtrapprox10^{32}$\lum) in the pulsar state is a common feature of all three known transitional MSPs (tMSPs; i.e., PSR J1023+0038, PSR J1227$-$4853, and M28I), it has been suggested by \cite{2014ApJ...795...72L} that $L_X\gtrapprox10^{32}$ is possibly a consequence of a stronger interaction between the pulsar and the companion, and therefore the higher X-ray luminosity could be a signature of a RB binary developing a strong accretion for the transition. {One possibility is that the companion of a pre-transition (to the LMXB state) system has a stronger wind {(i.e, a stronger inflow to the pulsar; see \citealt{2014ApJ...785..131T} and \citealt{2014ApJ...797..111L} for the interpretation of a varying stellar wind as the transition trigger for PSR J1023+0038)}, which powers a stronger intrabinary shock X-ray emission.} Based on the X-ray luminosity, two bright systems, PSR J2215+5135 ($L_X=1.3\times10^{32}$\lum) and PSR J1723-2837 ($L_X=2.4\times10^{32}$\lum; see Table \ref{tab:rbs} for their $\gamma$/X-ray properties), have been suggested by \cite{2014ApJ...795...72L} to be {potential targets for state transitions in the near future}. 3FGL J0212.1+5320 could be the third member of the group. In addition, we also examined the X-ray-to-$\gamma$-ray flux ratios of some known RBs and found that the flux ratios of the tMSPs (i.e., $\gg 1$\%) are significantly larger than that of the ``normal'' RBs (i.e., $\lesssim1$\%). 3FGL J0212.1+5320 has a ratio of $7.9$\% that is consistent with the tMSP ones. {One of the two prospective tMSP candidates, PSR J1723$-$2837, also has a large ratio of 13\% (Table \ref{tab:rbs}). }

{Certainly, the speculation is not mature and should not be taken conclusively. However, it is still worth paying attention to the X-ray activity of 3FGL J0212.1+5320 for any future transition.} Even if it is not exhibiting any transition in the near future, 3FGL J0212.1+5320 could be one of the brightest RBs in X-rays and certainly is one of the best sources {for studying} the X-ray emissions of RBs. 

No previous attempt of radio pulsation blind search for 3FGL J0212.1+5320 has been found in the literature \citep{2011ApJ...727L..16R,2012MNRAS.422.1294G,2015ApJ...810...85C}. In fact, the system is likely radio-faint as no radio counterpart can be found in the 1.4 GHz NRAO/VLA Sky Survey (NVSS), of which the detection limit is $\sim2.5$ mJy {(\citealt{1998AJ....115.1693C}; Note: most of the radio MSPs found by targeting \textit{Fermi}-LAT sources have flux densities much lower than 2.5 mJy at 1.4 GHz; \citealt{2012arXiv1205.3089R})}. Nevertheless, a GBT observation is being planned for searching for radio coherent pulsations. {Hopefully, this extreme RB MSP (i.e., high X-ray luminosity, bright optical companion, long orbital period, and potentially young age) can be confirmed soon. }

{After the submission of this paper, we became aware of a similar work by \cite{2016arXiv160902232L}, in which results including the measured orbital period, the radial velocity curve of the companion, the \textit{Chandra} spectral analysis, and the redback MSP nature interpretation are consistent with ours. In particular, they have sampled a much better radial velocity curve, which would be very helpful in searching the radio/$\gamma$-ray pulsations in the future. }

\begin{acknowledgements}

Support for this work was partially provided by the National Aeronautics and Space Administration through \textit{Chandra} Award Number DD5-16078X issued by the \textit{Chandra} X-ray Observatory Center, which is operated by the Smithsonian Astrophysical Observatory for and on behalf of the National Aeronautics Space Administration under contract NAS8-03060. 
Support from NASA grant NNX15AU83G is gratefully acknowledged. AKHK and XH are supported by the Ministry of Science and Technology of Taiwan through grant 103-2628-M-007-003-MY3 and 104-2811-M-007-059. 
XH is also supported by National Natural Science Foundation of China through grant 11503078. 
Jirong Mao is supported by the Hundred Talent Program of Chinese Academy of Sciences, the Key Research Program of Chinese Academy of Sciences (grant No. KJZD-EW-M06), and the Introducing Oversea Talent Plan of Yunnan Province. 
J.S.~acknowledges support from a Packard Fellowship. 
The \textit{Lulin} Observatory is operated by the Graduate Institute of Astronomy in National Central University, Taiwan. 
We acknowledge the support of the staff of the \textit{Lijiang} 2.4m telescope. Funding for the telescope has been provided by Chinese Academy of Sciences and the People's Government of Yunnan Province. 
The scientific results reported in this article are based in part on data obtained from the \textit{Chandra} Data Archive. 
We acknowledge the use of public data from the \textit{Swift} data archive. 
This publication makes use of data products from the \textit{Wide-field Infrared Survey Explorer}, which is a joint project of the University of California, Los Angeles, and the Jet Propulsion Laboratory/California Institute of Technology, funded by the National Aeronautics and Space Administration. 
\end{acknowledgements}

\bibliography{j0212}

\begin{thebibliography}{59}
\expandafter\ifx\csname natexlab\endcsname\relax\def\natexlab#1{#1}\fi

\bibitem[{{Abdo} {et~al.}(2010){Abdo}, {Ackermann}, {Ajello}, {Allafort},
  {Antolini}, {Atwood}, {Axelsson}, {Baldini}, {Ballet}, {Barbiellini}, \&
  et~al.}]{2010ApJS..188..405A}
{Abdo}, A.~A., {Ackermann}, M., {Ajello}, M., {et~al.} 2010, \apjs, 188, 405

\bibitem[{{Abdo} {et~al.}(2013){Abdo}, {Ajello}, {Allafort}, {Baldini},
  {Ballet}, {Barbiellini}, {Baring}, {Bastieri}, {Belfiore}, {Bellazzini}, \&
  et~al.}]{2013ApJS..208...17A}
{Abdo}, A.~A., {Ajello}, M., {Allafort}, A., {et~al.} 2013, \apjs, 208, 17

\bibitem[{{Acero} {et~al.}(2015){Acero}, {Ackermann}, {Ajello}, {Albert},
  {Atwood}, {Axelsson}, {Baldini}, {Ballet}, {Barbiellini}, {Bastieri},
  {Belfiore}, {Bellazzini}, {Bissaldi}, {Blandford}, {Bloom}, {Bogart},
  {Bonino}, {Bottacini}, {Bregeon}, {Britto}, {Bruel}, {Buehler}, {Burnett},
  {Buson}, {Caliandro}, {Cameron}, {Caputo}, {Caragiulo}, {Caraveo},
  {Casandjian}, {Cavazzuti}, {Charles}, {Chaves}, {Chekhtman}, {Cheung},
  {Chiang}, {Chiaro}, {Ciprini}, {Claus}, {Cohen-Tanugi}, {Cominsky}, {Conrad},
  {Cutini}, {D'Ammando}, {de Angelis}, {DeKlotz}, {de Palma}, {Desiante},
  {Digel}, {Di Venere}, {Drell}, {Dubois}, {Dumora}, {Favuzzi}, {Fegan},
  {Ferrara}, {Finke}, {Franckowiak}, {Fukazawa}, {Funk}, {Fusco}, {Gargano},
  {Gasparrini}, {Giebels}, {Giglietto}, {Giommi}, {Giordano}, {Giroletti},
  {Glanzman}, {Godfrey}, {Grenier}, {Grondin}, {Grove}, {Guillemot}, {Guiriec},
  {Hadasch}, {Harding}, {Hays}, {Hewitt}, {Hill}, {Horan}, {Iafrate}, {Jogler},
  {J{\'o}hannesson}, {Johnson}, {Johnson}, {Johnson}, {Johnson}, {Kamae},
  {Kataoka}, {Katsuta}, {Kuss}, {La Mura}, {Landriu}, {Larsson}, {Latronico},
  {Lemoine-Goumard}, {Li}, {Li}, {Longo}, {Loparco}, {Lott}, {Lovellette},
  {Lubrano}, {Madejski}, {Massaro}, {Mayer}, {Mazziotta}, {McEnery},
  {Michelson}, {Mirabal}, {Mizuno}, {Moiseev}, {Mongelli}, {Monzani},
  {Morselli}, {Moskalenko}, {Murgia}, {Nuss}, {Ohno}, {Ohsugi}, {Omodei},
  {Orienti}, {Orlando}, {Ormes}, {Paneque}, {Panetta}, {Perkins},
  {Pesce-Rollins}, {Piron}, {Pivato}, {Porter}, {Racusin}, {Rando}, {Razzano},
  {Razzaque}, {Reimer}, {Reimer}, {Reposeur}, {Rochester}, {Romani},
  {Salvetti}, {S{\'a}nchez-Conde}, {Saz Parkinson}, {Schulz}, {Siskind},
  {Smith}, {Spada}, {Spandre}, {Spinelli}, {Stephens}, {Strong}, {Suson},
  {Takahashi}, {Takahashi}, {Tanaka}, {Thayer}, {Thayer}, {Thompson},
  {Tibaldo}, {Tibolla}, {Torres}, {Torresi}, {Tosti}, {Troja}, {Van Klaveren},
  {Vianello}, {Winer}, {Wood}, {Wood}, {Zimmer}, \& {Fermi-LAT
  Collaboration}}]{2015ApJS..218...23A}
{Acero}, F., {Ackermann}, M., {Ajello}, M., {et~al.} 2015, \apjs, 218, 23

\bibitem[{{Alpar} {et~al.}(1982){Alpar}, {Cheng}, {Ruderman}, \&
  {Shaham}}]{1982Natur.300..728A}
{Alpar}, M.~A., {Cheng}, A.~F., {Ruderman}, M.~A., \& {Shaham}, J. 1982, \nat,
  300, 728

\bibitem[{{Archibald} {et~al.}(2009){Archibald}, {Stairs}, {Ransom}, {Kaspi},
  {Kondratiev}, {Lorimer}, {McLaughlin}, {Boyles}, {Hessels}, {Lynch}, {van
  Leeuwen}, {Roberts}, {Jenet}, {Champion}, {Rosen}, {Barlow}, {Dunlap}, \&
  {Remillard}}]{2009Sci...324.1411A}
{Archibald}, A.~M., {Stairs}, I.~H., {Ransom}, S.~M., {et~al.} 2009, Science,
  324, 1411

\bibitem[{{Bellm} {et~al.}(2016){Bellm}, {Kaplan}, {Breton}, {Phinney},
  {Bhalerao}, {Camilo}, {Dahal}, {Djorgovski}, {Drake}, {Hessels}, {Laher},
  {Levitan}, {Lewis}, {Mahabal}, {Ofek}, {Prince}, {Ransom}, {Roberts},
  {Russell}, {Sesar}, {Surace}, \& {Tang}}]{2016ApJ...816...74B}
{Bellm}, E.~C., {Kaplan}, D.~L., {Breton}, R.~P., {et~al.} 2016, \apj, 816, 74

\bibitem[{{Binney} \& {Merrifield}(1998)}]{1998gaas.book.....B}
{Binney}, J., \& {Merrifield}, M. 1998, {Galactic Astronomy}

\bibitem[{{Bogdanov} {et~al.}(2011){Bogdanov}, {Archibald}, {Hessels}, {Kaspi},
  {Lorimer}, {McLaughlin}, {Ransom}, \& {Stairs}}]{2011ApJ...742...97B}
{Bogdanov}, S., {Archibald}, A.~M., {Hessels}, J.~W.~T., {et~al.} 2011, \apj,
  742, 97

\bibitem[{{Camilo} {et~al.}(2015){Camilo}, {Kerr}, {Ray}, {Ransom},
  {Sarkissian}, {Cromartie}, {Johnston}, {Reynolds}, {Wolff}, {Freire},
  {Bhattacharyya}, {Ferrara}, {Keith}, {Michelson}, {Saz Parkinson}, \&
  {Wood}}]{2015ApJ...810...85C}
{Camilo}, F., {Kerr}, M., {Ray}, P.~S., {et~al.} 2015, \apj, 810, 85

\bibitem[{{Camilo} {et~al.}(2016){Camilo}, {Reynolds}, {Ransom}, {Halpern},
  {Bogdanov}, {Kerr}, {Ray}, {Cordes}, {Sarkissian}, {Barr}, \&
  {Ferrara}}]{2016ApJ...820....6C}
{Camilo}, F., {Reynolds}, J.~E., {Ransom}, S.~M., {et~al.} 2016, \apj, 820, 6

\bibitem[{{Cardelli} {et~al.}(1989){Cardelli}, {Clayton}, \&
  {Mathis}}]{1989ApJ...345..245C}
{Cardelli}, J.~A., {Clayton}, G.~C., \& {Mathis}, J.~S. 1989, \apj, 345, 245

\bibitem[{{Chen} {et~al.}(2013){Chen}, {Chen}, {Tauris}, \&
  {Han}}]{2013ApJ...775...27C}
{Chen}, H.-L., {Chen}, X., {Tauris}, T.~M., \& {Han}, Z. 2013, \apj, 775, 27

\bibitem[{{Cheng} {et~al.}(1986){Cheng}, {Ho}, \&
  {Ruderman}}]{1986ApJ...300..500C}
{Cheng}, K.~S., {Ho}, C., \& {Ruderman}, M. 1986, \apj, 300, 500

\bibitem[{{Condon} {et~al.}(1998){Condon}, {Cotton}, {Greisen}, {Yin},
  {Perley}, {Taylor}, \& {Broderick}}]{1998AJ....115.1693C}
{Condon}, J.~J., {Cotton}, W.~D., {Greisen}, E.~W., {et~al.} 1998, \aj, 115,
  1693

\bibitem[{{Deneva} {et~al.}(2016){Deneva}, {Ray}, {Camilo}, {Halpern}, {Wood},
  {Cromartie}, {Ferrara}, {Kerr}, {Ransom}, {Wolff}, {Chambers}, \&
  {Magnier}}]{2016ApJ...823..105D}
{Deneva}, J.~S., {Ray}, P.~S., {Camilo}, F., {et~al.} 2016, \apj, 823, 105

\bibitem[{{Evans} {et~al.}(2014){Evans}, {Osborne}, {Beardmore}, {Page},
  {Willingale}, {Mountford}, {Pagani}, {Burrows}, {Kennea}, {Perri},
  {Tagliaferri}, \& {Gehrels}}]{2014ApJS..210....8E}
{Evans}, P.~A., {Osborne}, J.~P., {Beardmore}, A.~P., {et~al.} 2014, \apjs,
  210, 8

\bibitem[{{Fan} {et~al.}(2015){Fan}, {Bai}, {Zhang}, {Wang}, {Chang}, {Xin}, \&
  {Zhang}}]{2015RAA....15..918F}
{Fan}, Y.-F., {Bai}, J.-M., {Zhang}, J.-J., {et~al.} 2015, Research in
  Astronomy and Astrophysics, 15, 918

\bibitem[{{Freire} {et~al.}(2011){Freire}, {Abdo}, {Ajello}, {Allafort},
  {Ballet}, {Barbiellini}, {Bastieri}, {Bechtol}, {Bellazzini}, {Blandford},
  {Bloom}, {Bonamente}, {Borgland}, {Brigida}, {Bruel}, {Buehler}, {Buson},
  {Caliandro}, {Cameron}, {Camilo}, {Caraveo}, {Cecchi}, {{\c C}elik},
  {Charles}, {Chekhtman}, {Cheung}, {Chiang}, {Ciprini}, {Claus}, {Cognard},
  {Cohen-Tanugi}, {Cominsky}, {de Palma}, {Dermer}, {do Couto e Silva},
  {Dormody}, {Drell}, {Dubois}, {Dumora}, {Espinoza}, {Favuzzi}, {Fegan},
  {Ferrara}, {Focke}, {Fortin}, {Fukazawa}, {Fusco}, {Gargano}, {Gasparrini},
  {Gehrels}, {Germani}, {Giglietto}, {Giordano}, {Giroletti}, {Glanzman},
  {Godfrey}, {Grenier}, {Grondin}, {Grove}, {Guillemot}, {Guiriec}, {Hadasch},
  {Harding}, {J{\'o}hannesson}, {Johnson}, {Johnson}, {Johnston}, {Katagiri},
  {Kataoka}, {Keith}, {Kerr}, {Kn{\"o}dlseder}, {Kramer}, {Kuss}, {Lande},
  {Latronico}, {Lee}, {Lemoine-Goumard}, {Longo}, {Loparco}, {Lovellette},
  {Lubrano}, {Lyne}, {Manchester}, {Marelli}, {Mazziotta}, {McEnery},
  {Michelson}, {Mizuno}, {Moiseev}, {Monte}, {Monzani}, {Morselli},
  {Moskalenko}, {Murgia}, {Nakamori}, {Nolan}, {Norris}, {Nuss}, {Ohsugi},
  {Okumura}, {Omodei}, {Orlando}, {Ozaki}, {Paneque}, {Parent},
  {Pesce-Rollins}, {Pierbattista}, {Piron}, {Porter}, {Rain{\`o}}, {Ransom},
  {Ray}, {Reimer}, {Reimer}, {Reposeur}, {Ritz}, {Romani}, {Roth},
  {Sadrozinski}, {Saz Parkinson}, {Shannon}, {Siskind}, {Smith}, {Spinelli},
  {Stappers}, {Suson}, {Takahashi}, {Tanaka}, {Tauris}, {Thayer}, {Theureau},
  {Thompson}, {Thorsett}, {Tibaldo}, {Torres}, {Tosti}, {Troja},
  {Vandenbroucke}, {Van Etten}, {Vasileiou}, {Venter}, {Vianello}, {Vilchez},
  {Vitale}, {Waite}, {Wang}, {Wood}, {Yang}, {Ziegler}, \&
  {Zimmer}}]{2011Sci...334.1107F}
{Freire}, P.~C.~C., {Abdo}, A.~A., {Ajello}, M., {et~al.} 2011, Science, 334,
  1107

\bibitem[{{Guillemot} {et~al.}(2012){Guillemot}, {Freire}, {Cognard},
  {Johnson}, {Takahashi}, {Kataoka}, {Desvignes}, {Camilo}, {Ferrara},
  {Harding}, {Janssen}, {Keith}, {Kerr}, {Kramer}, {Parent}, {Ransom}, {Ray},
  {Saz Parkinson}, {Smith}, {Stappers}, \& {Theureau}}]{2012MNRAS.422.1294G}
{Guillemot}, L., {Freire}, P.~C.~C., {Cognard}, I., {et~al.} 2012, \mnras, 422,
  1294

\bibitem[{{Hui} {et~al.}(2015{\natexlab{a}}){Hui}, {Hu}, {Park}, {Takata},
  {Li}, {Tam}, {Lin}, {Kong}, {Cheng}, \& {Kim}}]{2015ApJ...801L..27H}
{Hui}, C.~Y., {Hu}, C.~P., {Park}, S.~M., {et~al.} 2015{\natexlab{a}}, \apjl,
  801, L27

\bibitem[{{Hui} {et~al.}(2015{\natexlab{b}}){Hui}, {Park}, {Hu}, {Lin}, {Li},
  {Kong}, {Tam}, {Takata}, {Cheng}, {Jin}, {Yen}, \&
  {Kim}}]{2015ApJ...809...68H}
{Hui}, C.~Y., {Park}, S.~M., {Hu}, C.~P., {et~al.} 2015{\natexlab{b}}, \apj,
  809, 68

\bibitem[{{Joss} \& {Rappaport}(1984)}]{1984ARA&A..22..537J}
{Joss}, P.~C., \& {Rappaport}, S.~A. 1984, \araa, 22, 537

\bibitem[{{Kalberla} {et~al.}(2005){Kalberla}, {Burton}, {Hartmann}, {Arnal},
  {Bajaja}, {Morras}, \& {P{\"o}ppel}}]{2005A&A...440..775K}
{Kalberla}, P.~M.~W., {Burton}, W.~B., {Hartmann}, D., {et~al.} 2005, \aap,
  440, 775

\bibitem[{{Kong} {et~al.}(2012){Kong}, {Huang}, {Cheng}, {Takata}, {Yatsu},
  {Cheung}, {Donato}, {Lin}, {Kataoka}, {Takahashi}, {Maeda}, {Hui}, \&
  {Tam}}]{2012ApJ...747L...3K}
{Kong}, A.~K.~H., {Huang}, R.~H.~H., {Cheng}, K.~S., {et~al.} 2012, \apjl, 747,
  L3

\bibitem[{{Kong} {et~al.}(2014){Kong}, {Jin}, {Yen}, {Hu}, {Hui}, {Tam},
  {Takata}, {Lin}, {Cheng}, {Park}, \& {Kim}}]{2014ApJ...794L..22K}
{Kong}, A.~K.~H., {Jin}, R., {Yen}, T.-C., {et~al.} 2014, \apjl, 794, L22

\bibitem[{{Li} {et~al.}(2014){Li}, {Kong}, {Takata}, {Cheng}, {Tam}, {Hui}, \&
  {Jin}}]{2014ApJ...797..111L}
{Li}, K.~L., {Kong}, A.~K.~H., {Takata}, J., {et~al.} 2014, \apj, 797, 111

\bibitem[{{Linares}(2014)}]{2014ApJ...795...72L}
{Linares}, M. 2014, \apj, 795, 72

\bibitem[{{Linares} {et~al.}(2016){Linares}, {Miles-P{\'a}ez},
  {Rodr{\'{\i}}guez-Gil}, {Shahbaz}, {Casares}, {Fari{\~n}a}, \&
  {Karjalainen}}]{2016arXiv160902232L}
{Linares}, M., {Miles-P{\'a}ez}, P., {Rodr{\'{\i}}guez-Gil}, P., {et~al.} 2016,
  (MNRAS Submitted)

\bibitem[{{Mirabal} {et~al.}(2016){Mirabal}, {Charles}, {Ferrara}, {Gonthier},
  {Harding}, {S{\'a}nchez-Conde}, \& {Thompson}}]{2016ApJ...825...69M}
{Mirabal}, N., {Charles}, E., {Ferrara}, E.~C., {et~al.} 2016, \apj, 825, 69

\bibitem[{{Monet} {et~al.}(2003){Monet}, {Levine}, {Canzian}, {Ables}, {Bird},
  {Dahn}, {Guetter}, {Harris}, {Henden}, {Leggett}, {Levison}, {Luginbuhl},
  {Martini}, {Monet}, {Munn}, {Pier}, {Rhodes}, {Riepe}, {Sell}, {Stone},
  {Vrba}, {Walker}, {Westerhout}, {Brucato}, {Reid}, {Schoening}, {Hartley},
  {Read}, \& {Tritton}}]{2003AJ....125..984M}
{Monet}, D.~G., {Levine}, S.~E., {Canzian}, B., {et~al.} 2003, \aj, 125, 984

\bibitem[{{Munari} {et~al.}(2005){Munari}, {Sordo}, {Castelli}, \&
  {Zwitter}}]{2005A&A...442.1127M}
{Munari}, U., {Sordo}, R., {Castelli}, F., \& {Zwitter}, T. 2005, \aap, 442,
  1127

\bibitem[{{Muslimov} \& {Harding}(2003)}]{2003ApJ...588..430M}
{Muslimov}, A.~G., \& {Harding}, A.~K. 2003, \apj, 588, 430

\bibitem[{{Oke}(1990)}]{1990AJ.....99.1621O}
{Oke}, J.~B. 1990, \aj, 99, 1621

\bibitem[{{Orosz} \& {Hauschildt}(2000)}]{2000A&A...364..265O}
{Orosz}, J.~A., \& {Hauschildt}, P.~H. 2000, \aap, 364, 265

\bibitem[{{Papitto} \& {Torres}(2015)}]{2015ApJ...807...33P}
{Papitto}, A., \& {Torres}, D.~F. 2015, \apj, 807, 33

\bibitem[{{Papitto} {et~al.}(2013){Papitto}, {Ferrigno}, {Bozzo}, {Rea},
  {Pavan}, {Burderi}, {Burgay}, {Campana}, {di Salvo}, {Falanga},
  {Filipovi{\'c}}, {Freire}, {Hessels}, {Possenti}, {Ransom}, {Riggio},
  {Romano}, {Sarkissian}, {Stairs}, {Stella}, {Torres}, {Wieringa}, \&
  {Wong}}]{2013Natur.501..517P}
{Papitto}, A., {Ferrigno}, C., {Bozzo}, E., {et~al.} 2013, \nat, 501, 517

\bibitem[{{Patruno} {et~al.}(2014){Patruno}, {Archibald}, {Hessels},
  {Bogdanov}, {Stappers}, {Bassa}, {Janssen}, {Kaspi}, {Tendulkar}, \&
  {Lyne}}]{2014ApJ...781L...3P}
{Patruno}, A., {Archibald}, A.~M., {Hessels}, J.~W.~T., {et~al.} 2014, \apjl,
  781, L3

\bibitem[{{Pletsch} \& {Clark}(2015)}]{2015ApJ...807...18P}
{Pletsch}, H.~J., \& {Clark}, C.~J. 2015, \apj, 807, 18

\bibitem[{{Pletsch} {et~al.}(2012){Pletsch}, {Guillemot}, {Fehrmann}, {Allen},
  {Kramer}, {Aulbert}, {Ackermann}, {Ajello}, {de Angelis}, {Atwood},
  {Baldini}, {Ballet}, {Barbiellini}, {Bastieri}, {Bechtol}, {Bellazzini},
  {Borgland}, {Bottacini}, {Brandt}, {Bregeon}, {Brigida}, {Bruel}, {Buehler},
  {Buson}, {Caliandro}, {Cameron}, {Caraveo}, {Casandjian}, {Cecchi}, {{\c
  C}elik}, {Charles}, {Chaves}, {Cheung}, {Chiang}, {Ciprini}, {Claus},
  {Cohen-Tanugi}, {Conrad}, {Cutini}, {D'Ammando}, {Dermer}, {Digel}, {Drell},
  {Drlica-Wagner}, {Dubois}, {Dumora}, {Favuzzi}, {Ferrara}, {Franckowiak},
  {Fukazawa}, {Fusco}, {Gargano}, {Gehrels}, {Germani}, {Giglietto},
  {Giordano}, {Giroletti}, {Godfrey}, {Grenier}, {Grondin}, {Grove}, {Guiriec},
  {Hadasch}, {Hanabata}, {Harding}, {den Hartog}, {Hayashida}, {Hays}, {Hill},
  {Hou}, {Hughes}, {J{\'o}hannesson}, {Jackson}, {Jogler}, {Johnson},
  {Johnson}, {Kataoka}, {Kerr}, {Kn{\"o}dlseder}, {Kuss}, {Lande}, {Larsson},
  {Latronico}, {Lemoine-Goumard}, {Longo}, {Loparco}, {Lovellette}, {Lubrano},
  {Massaro}, {Mayer}, {Mazziotta}, {McEnery}, {Mehault}, {Michelson},
  {Mitthumsiri}, {Mizuno}, {Monzani}, {Morselli}, {Moskalenko}, {Murgia},
  {Nakamori}, {Nemmen}, {Nuss}, {Ohno}, {Ohsugi}, {Omodei}, {Orienti},
  {Orlando}, {de Palma}, {Paneque}, {Perkins}, {Piron}, {Pivato}, {Porter},
  {Rain{\`o}}, {Rando}, {Ray}, {Razzano}, {Reimer}, {Reimer}, {Reposeur},
  {Ritz}, {Romani}, {Romoli}, {Sanchez}, {Parkinson}, {Schulz}, {Sgr{\`o}}, {do
  Couto e Silva}, {Siskind}, {Smith}, {Spandre}, {Spinelli}, {Suson},
  {Takahashi}, {Tanaka}, {Thayer}, {Thayer}, {Thompson}, {Tibaldo},
  {Tinivella}, {Troja}, {Usher}, {Vandenbroucke}, {Vasileiou}, {Vianello},
  {Vitale}, {Waite}, {Winer}, {Wood}, {Wood}, {Yang}, \&
  {Zimmer}}]{2012Sci...338.1314P}
{Pletsch}, H.~J., {Guillemot}, L., {Fehrmann}, H., {et~al.} 2012, Science, 338,
  1314

\bibitem[{{Ransom} {et~al.}(2011){Ransom}, {Ray}, {Camilo}, {Roberts}, {{\c
  C}elik}, {Wolff}, {Cheung}, {Kerr}, {Pennucci}, {DeCesar}, {Cognard}, {Lyne},
  {Stappers}, {Freire}, {Grove}, {Abdo}, {Desvignes}, {Donato}, {Ferrara},
  {Gehrels}, {Guillemot}, {Gwon}, {Harding}, {Johnston}, {Keith}, {Kramer},
  {Michelson}, {Parent}, {Saz Parkinson}, {Romani}, {Smith}, {Theureau},
  {Thompson}, {Weltevrede}, {Wood}, \& {Ziegler}}]{2011ApJ...727L..16R}
{Ransom}, S.~M., {Ray}, P.~S., {Camilo}, F., {et~al.} 2011, \apjl, 727, L16

\bibitem[{{Ray} {et~al.}(2012){Ray}, {Abdo}, {Parent}, {Bhattacharya},
  {Bhattacharyya}, {Camilo}, {Cognard}, {Theureau}, {Ferrara}, {Harding},
  {Thompson}, {Freire}, {Guillemot}, {Gupta}, {Roy}, {Hessels}, {Johnston},
  {Keith}, {Shannon}, {Kerr}, {Michelson}, {Romani}, {Kramer}, {McLaughlin},
  {Ransom}, {Roberts}, {Saz Parkinson}, {Ziegler}, {Smith}, {Stappers},
  {Weltevrede}, \& {Wood}}]{2012arXiv1205.3089R}
{Ray}, P.~S., {Abdo}, A.~A., {Parent}, D., {et~al.} 2012, 2011 Fermi Symposium
  proceedings - eConf C110509

\bibitem[{{Roy} {et~al.}(2015){Roy}, {Ray}, {Bhattacharyya}, {Stappers},
  {Chengalur}, {Deneva}, {Camilo}, {Johnson}, {Wolff}, {Hessels}, {Bassa},
  {Keane}, {Ferrara}, {Harding}, \& {Wood}}]{2015ApJ...800L..12R}
{Roy}, J., {Ray}, P.~S., {Bhattacharyya}, B., {et~al.} 2015, \apjl, 800, L12

\bibitem[{{Ruderman} \& {Sutherland}(1975)}]{1975ApJ...196...51R}
{Ruderman}, M.~A., \& {Sutherland}, P.~G. 1975, \apj, 196, 51

\bibitem[{{Saz Parkinson} {et~al.}(2016){Saz Parkinson}, {Xu}, {Yu},
  {Salvetti}, {Marelli}, \& {Falcone}}]{2016ApJ...820....8S}
{Saz Parkinson}, P.~M., {Xu}, H., {Yu}, P.~L.~H., {et~al.} 2016, \apj, 820, 8

\bibitem[{{Schlafly} \& {Finkbeiner}(2011)}]{2011ApJ...737..103S}
{Schlafly}, E.~F., \& {Finkbeiner}, D.~P. 2011, \apj, 737, 103

\bibitem[{{Sing}(2010)}]{2010A&A...510A..21S}
{Sing}, D.~K. 2010, \aap, 510, A21

\bibitem[{{Skrutskie} {et~al.}(2006){Skrutskie}, {Cutri}, {Stiening},
  {Weinberg}, {Schneider}, {Carpenter}, {Beichman}, {Capps}, {Chester},
  {Elias}, {Huchra}, {Liebert}, {Lonsdale}, {Monet}, {Price}, {Seitzer},
  {Jarrett}, {Kirkpatrick}, {Gizis}, {Howard}, {Evans}, {Fowler}, {Fullmer},
  {Hurt}, {Light}, {Kopan}, {Marsh}, {McCallon}, {Tam}, {Van Dyk}, \&
  {Wheelock}}]{2006AJ....131.1163S}
{Skrutskie}, M.~F., {Cutri}, R.~M., {Stiening}, R., {et~al.} 2006, \aj, 131,
  1163

\bibitem[{{Strader} {et~al.}(2015){Strader}, {Chomiuk}, {Cheung}, {Sand},
  {Donato}, {Corbet}, {Koeppe}, {Edwards}, {Stevens}, {Petrov}, {Salinas},
  {Peacock}, {Finzell}, {Reichart}, \& {Haislip}}]{2015ApJ...804L..12S}
{Strader}, J., {Chomiuk}, L., {Cheung}, C.~C., {et~al.} 2015, \apjl, 804, L12

\bibitem[{{Stroh} \& {Falcone}(2013)}]{2013ApJS..207...28S}
{Stroh}, M.~C., \& {Falcone}, A.~D. 2013, \apjs, 207, 28

\bibitem[{{Takata} {et~al.}(2014){Takata}, {Li}, {Leung}, {Kong}, {Tam}, {Hui},
  {Wu}, {Xing}, {Cao}, {Tang}, {Wang}, \& {Cheng}}]{2014ApJ...785..131T}
{Takata}, J., {Li}, K.~L., {Leung}, G.~C.~K., {et~al.} 2014, \apj, 785, 131

\bibitem[{{Tam} {et~al.}(2010){Tam}, {Hui}, {Huang}, {Kong}, {Takata}, {Lin},
  {Yang}, {Cheng}, \& {Taam}}]{2010ApJ...724L.207T}
{Tam}, P.~H.~T., {Hui}, C.~Y., {Huang}, R.~H.~H., {et~al.} 2010, \apjl, 724,
  L207

\bibitem[{{Tauris}(2011)}]{2011ASPC..447..285T}
{Tauris}, T.~M. 2011, in Astronomical Society of the Pacific Conference Series,
  Vol. 447, Evolution of Compact Binaries, ed. L.~{Schmidtobreick}, M.~R.
  {Schreiber}, \& C.~{Tappert}, 285

\bibitem[{{Tendulkar} {et~al.}(2014){Tendulkar}, {Yang}, {An}, {Kaspi},
  {Archibald}, {Bassa}, {Bellm}, {Bogdanov}, {Harrison}, {Hessels}, {Janssen},
  {Lyne}, {Patruno}, {Stappers}, {Stern}, {Tomsick}, {Boggs}, {Chakrabarty},
  {Christensen}, {Craig}, {Hailey}, \& {Zhang}}]{2014ApJ...791...77T}
{Tendulkar}, S.~P., {Yang}, C., {An}, H., {et~al.} 2014, \apj, 791, 77

\bibitem[{{van den Heuvel} \& {van Paradijs}(1988)}]{1988Natur.334..227V}
{van den Heuvel}, E.~P.~J., \& {van Paradijs}, J. 1988, \nat, 334, 227

\bibitem[{{van Hamme}(1993)}]{1993AJ....106.2096V}
{van Hamme}, W. 1993, \aj, 106, 2096

\bibitem[{{Wright} {et~al.}(2010){Wright}, {Eisenhardt}, {Mainzer}, {Ressler},
  {Cutri}, {Jarrett}, {Kirkpatrick}, {Padgett}, {McMillan}, {Skrutskie},
  {Stanford}, {Cohen}, {Walker}, {Mather}, {Leisawitz}, {Gautier}, {McLean},
  {Benford}, {Lonsdale}, {Blain}, {Mendez}, {Irace}, {Duval}, {Liu}, {Royer},
  {Heinrichsen}, {Howard}, {Shannon}, {Kendall}, {Walsh}, {Larsen}, {Cardon},
  {Schick}, {Schwalm}, {Abid}, {Fabinsky}, {Naes}, \&
  {Tsai}}]{2010AJ....140.1868W}
{Wright}, E.~L., {Eisenhardt}, P.~R.~M., {Mainzer}, A.~K., {et~al.} 2010, \aj,
  140, 1868

\bibitem[{{Xing} \& {Wang}(2015)}]{2015ApJ...808...17X}
{Xing}, Y., \& {Wang}, Z. 2015, \apj, 808, 17

\bibitem[{{Zacharias} {et~al.}(2013){Zacharias}, {Finch}, {Girard}, {Henden},
  {Bartlett}, {Monet}, \& {Zacharias}}]{2013AJ....145...44Z}
{Zacharias}, N., {Finch}, C.~T., {Girard}, T.~M., {et~al.} 2013, \aj, 145, 44

\bibitem[{{Zahn}(1977)}]{1977A&A....57..383Z}
{Zahn}, J.-P. 1977, \aap, 57, 383

\end{thebibliography}
\end{document}